\documentclass[sn-basic]{sn-jnl}
\usepackage{graphicx}%
\usepackage{multirow}%
\usepackage{amsmath,amssymb,amsfonts}%
\usepackage{amsthm}%
\usepackage{mathrsfs}%
\usepackage{xcolor}%
\usepackage{textcomp}%
\usepackage{manyfoot}%
\usepackage{booktabs}%
\usepackage{algorithm}%
\usepackage{algorithmicx}%
\usepackage{algpseudocode}%
\usepackage{listings}%
\theoremstyle{thmstyleone}%
\theoremstyle{thmstyletwo}%
\theoremstyle{thmstylethree}%

\newcommand*\m{\mathrm{m}}                     

\newcommand*\s{\mathrm{s}}                     

\newcommand*\kg{\mathrm{kg}}                   

\newcommand*\Del{\mathnormal{\Delta}}
\newcommand{\km}{ {\mathrm {km} }}
\newcommand{\Mpc}{ {\mathrm {Mpc} }}
\renewcommand{\vec}[1]{\mbox{\boldmath $#1$}}
\def\deg{\hbox{$^{\rm o}$}}

\usepackage{atbegshi}
\AtBeginDocument{\AtBeginShipoutNext{\AtBeginShipoutDiscard}}

\begin{document}
\setcounter{page}{0}

\title[Article Title]{Doppler, gravitational and cosmological redshifts}


\author[1]{\fnm{Klaus} \sur{Wilhelm}}
\email{wilhelm@mps.mpg.de}.
\equalcont{This author contributed equally to this work.}

\author*[2]{\fnm{Bhola N.} \sur{Dwivedi}}
\email{bholadwivedi@gmail.com}.
\equalcont{This author contributed equally to this work.}

\affil[1]{\orgdiv{Max-Planck-Institut f\"ur Son\-nen\-sy\-stem\-for\-schung
(MPS)}, \street{Justus-von-Liebig-Weg 3},
\city{37077 G\"ottingen},
\country{Germany}}

\affil[2]{\orgdiv{Dept. of Physics (Retd. Professor)},
\orgname{Indian Institute of Technology (BHU)}, \city{Varanasi-221005},
\country{India}}


\abstract{\citet{OrtIba} have presented
a "Generalized redshift formula" taking account of only energy
conservation considerations. Contrary to their claim, we emphasize
to invoke both energy and momentum considerations in order to
deduce all three types of redshift (Doppler, gravitational and
cosmological). We formulate our views on the three physical
effects in a consistent manner in addition to addressing the
lack of relevant references in Ref.~\citep{OrtIba}.}

\keywords{Redshift, Doppler effect, gravitational redshift,
cosmological redshift, Hubble tension}

\maketitle






\section{Introduction}\label{sec:Introduction}

In an interesting article in {\bf scientific} reports
\citet{OrtIba}
presented a "Generalized redshift formula".
The authors claim that they can deduce all three types
of redshift (Doppler, gravitational and cosmological) from energy
conservation considerations. The main objective of our work is to assert
that we ought to invoke both energy and momentum considerations to deduce
all three types of redshift.
We have also difficulties with some of the statements and the
lack of relevant references in Ref.~\citep{OrtIba}.
We do not want to discuss these problems point by point, but rather formulate
our views on the three physical effects as consistently as possible
in the next three sections.

\section{Doppler effect and aberration}
\label{sec:Doppler}

Einstein's statement \citep{Ein24} highlights the importance of the
Doppler effect \citep{Dop42} and the aberration first described
by \citet{Bra27}:
\begin{enumerate} \item[ ]
``Whatever will eventually be the theory of electromagnetic processes,
the D\,o\,p\,p\,l\,e\,r principle and the aberration law will continue to be
valid, [...]'' . \footnote{This and
other quotations by Einstein are
originally in German. Emphasis on D\,o\,p\,p\,l\,e\,r,  E\,n\,e\,r\,g\,i\,e
and I\,m\,p\,u\,l\,s (i.e., momentum)  by Einstein.}
\end{enumerate}


The Doppler effect in acoustic and optical configurations was discovered
by Doppler in 1842 and applied, among others, to double-star systems.\\
\citet{Bol43} immediately recognized the importance of the effect for
studies of the motion of astronomical objects.

Two other statements by \citet[][pp.~127 and 128]{Ein17}
\begin{enumerate} \item[ ]
``If a light beam hits a molecule and leads to an absorption or
emission of the radiation energy~$h\,\nu$ by an elementary process, this will
always be accompanied by a momentum transfer of
$h\,\nu/c$ to the molecule, [...].'' \\
``However one usually only considers the e\,n\,e\,r\,g\,y exchange without
taking the m\,o\,m\,e\,n\,t\,u\,m exchange into account.''
\end{enumerate}


indicate that for an understanding of the optical Doppler effect it is
essential to invoke both the momentum and energy conservation
principles.\footnote{$\nu$ is the frequency of the light.}
\footnote{The Planck constant (exact):
$h = 6.626\,070\,15 \times 10^{-34}$~J\,Hz$^{-1}$.}
\footnote{The speed of light
in vacuum (exact): $c_0 = 299\,792\,458~\m~\s^{-1}$. We write $c_0$ without
gravitational fields, otherwise $c$ or $C$.}
\footnote{These and other physical constants are taken from CODATA Internationally
recommended 2022 values, cf. \citet{BIPM_19}.}\\

The interaction of photons\footnote{\citet{Ein05a} used the
expressions ,,Energiequanten''
(energy quanta) and ,,Lichtquant'' (light quantum). The name ``photon'' was
later coined by \citet{Lew26}.} with particles (e.g., atoms, ions or
molecules) is governed by two
equations derived by \citet{Ein05a,Ein05c}.
They can be written in a modern format as
%
\begin{equation}
E_\nu = h\,\nu
\label{Equation_1}
\end{equation}
and
%
\begin{equation}
E_0 = m\, c_0^2~,
\label{Equation_2}
\end{equation}
where $E_\nu$ is the energy quantum of electromagnetic
radiation with a
frequency $\nu$ and
$E_0$ is the energy of the mass~$m$ at rest. Energy
and momentum of a free massive particle moving with a velocity~$\vec{v}$
relative to a reference frame~S are
%
\begin{equation}
E^2 = m^2\,c_0^4 + \vec{p}^{\,2}\,c_0^2
\label{Equation_3}
\end{equation}
and
%
\begin{equation}
\vec{p} = \vec{v}\,\frac{E}{c_0^2}~,
\label{Equation_4}
\end{equation}
where $E$ is the total energy, $\vec{p}$ the momentum
vector ($p = |\vec{p}|$),
and $m$ the ordinary mass, the same as in Newtonian
mechanics (cf. ``Letter from Albert Einstein to Lincoln Barnett'',
19 June 1948 \citep{Oku89,Oku09}).
With $\beta = v/c_0$, where $v = |\vec{v}| < c_0$, and the Lorentz factor
$\gamma = (1 - \beta^2)^{-1/2} \ge 1$ it is
\index{Lorentz factor}
%
\begin{equation}
E = \gamma\,m\,c_0^2 ~.
\label{Equation_5}
\end{equation}
The kinetic energy of the particle in an inertial system~S is
%
\begin{equation}
E_{\rm kin} = E - E_0 = m\,c_0^2\,(\gamma - 1).
\label{Equation_6}
\end{equation}
The mass is zero for photons\footnote{A zero mass follows from the
special theory of relativity (STR) and a speed of light in vacuum constant for all
frequencies. Various methods have been used to constrain the photon
mass to $m_\nu < 10^{-49}~\kg$, cf. \citep{Amsetal,GolNie}.}
and Eq.~(\ref{Equation_3}) reduces  to
%
\begin{equation}
E_\nu = p_\nu\,c_0
\label{Equation_7}
\end{equation}
in a region with a gravitational
potential~$U_0 = 0$
(for a definition see Eq.~(\ref{Equation_18}) in Section~\ref{sec:Gravitation}).

The relativistic Doppler effect can be
formulated as a direct consequence of the
Special Theory of Relativity (STR) \citep{Ein05b} and only depends on the
relative motion of the transmitter and receiver.
The aberration formula then
follows from the Lorentz transformations \citep{Lor95,Lor03},
cf. \citep[][p. 30]{WilFroe}.

\citet[][p.~105]{Fer32} deduced the non-relativistic optical Doppler effect
differently and stated explicitly that energy and momentum conservations
are important:
\begin{enumerate} \item[ ]
``The change of frequency of the light emitted from a moving source is very
simply explained by the wave theory of light. But it finds also a simple,
though apparently very different, explanation in the light-quantum theory;
it can be shown that the Doppler effect may be deduced from the conservation
of energy and momentum in the emission process.''
\end{enumerate}
We apply Fermi's idea and derive the Doppler effect and the
aberration formula also for relativistic cases. However, we consider
a photon source at rest in an inertial system~S and moving receivers in
systems~S$'$, rather than a moving transmitter as in Ref.~\citep{Fer32},
because its velocity cannot unambiguously be defined.\footnote{It must,
however, be emphasized that Fermi's idea works for all velocities
under the assumption on a general rest frame.}

\begin{figure}[t]
\centering
\includegraphics[width=\textwidth]{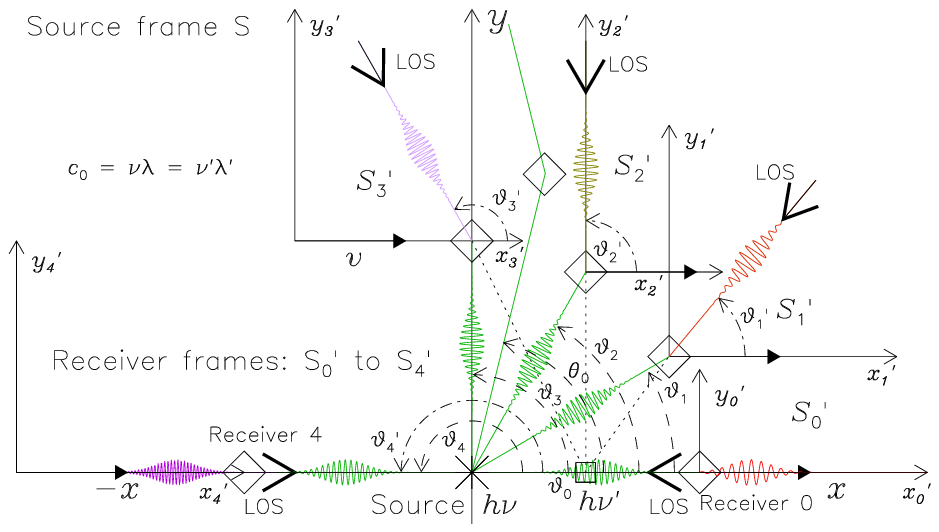}
\caption[]{\small
Photons with energy $h\,\nu$ are emitted under angles
$\vartheta_0$ to $\vartheta_4$ relative to the $+x$ coordinate axis
from a source at rest in the source frame~S.
They are sketched
as green wave packets. The receivers
are moving in systems~S$_0'$ to S$_4'$ with a velocity~$\vec{v}$
in the $+x$ direction
relative to the source frame S.
The Doppler-shifted wave packets are shown in red, blue and violet with
energies of $h\,\nu_0'$ to $h\,\nu_4'$,
respectively, deflected by aberration to
angles $\vartheta_0'$ to $\vartheta_4'$.
Since all detectors are shown at the same separation from the source,
their line of sight (LOS) directions point to the same position
displaced from the source (see dotted lines and
square symbol labelled $h\,\nu_0'$).
No shift occurs for $\theta_0$ (green line) deflected to $\theta_0'$
and no aberration for $\vartheta_0 = 0\deg$ and $\vartheta_4 = 180\deg$.
The relative wavelengths and the aberration angles are
to scale for $\beta= 0.5$.}
\label{fig:photon_1}
\end{figure}

Both laws can be directly derived from the momentum and energy
conservation principles:

A source with mass~$m$ at rest in frame S emits photons with an
energy of $\Del E = h\,\nu$ and momentum $\Del P = h\,\nu/c_0$ under angles
of  $\vartheta_0$ to $\vartheta_4$ relative to the $x$-axis of S.
This configuration is shown in Fig.~\ref{fig:photon_1}.
Assuming a large mass and applying the M\"ossbauer effect, cf. \citep{Moe58},
one can avoid any significant kinetic energy effects by recoil.\label{en:kin}
All receivers 0 to 4 are moving with a velocity $\vec{v}$ in the $+x$-direction.
It is also assumed that they have large masses.
In this approximation Fermi's classical treatment of the
Doppler effect can, without too many complications, be repeated for
relativistic cases. An important consequence of this approximation
is that the velocities of the emitter and the
receivers remain nearly constant although their energy contents and momenta
change. It is sufficient to determine the momentum changes.
The momentum component change in the $+y'$-direction in a receiver frame S$'$ is
%
\begin{equation}
\Del p'_y = \Del p_y =  \frac{h\,\nu}{c_0}\,\sin \vartheta.
\label{Equation_8}
\end{equation}
In the $x'$-direction two relativistic effects have to be considered:

1. The $x'$-component increases by the
Lorentz factor $\gamma$ with respect to \\
$\Del p_ x = h\,\nu\,\cos \vartheta/c_0$,
due to the relative motion between S$'$ and S
along this direction.

2. A portion of the absorbed energy $h\,\nu$ is required to
adjust the total momentum component of the receiver and has to be
taken into account during the excitation of a detector atom, cf.
Eqs.~(\ref{Equation_4}) and (\ref{Equation_5}).
The energy that is
available in this component leads to a momentum change of
%
\begin{equation}
\Del p'_x =
\gamma\,\frac{h\,\nu}{c_0}\,\cos \vartheta
- v\,\gamma\,\frac{h\,\nu}{c_0^2} =
\gamma\,\frac{h\,\nu}{c_0}\,(\cos \vartheta - \beta)~.
\label{Equation_9}
\end{equation}
The square of the momentum change thus is
%
\begin{eqnarray}
(\Del P')^2 = (\Del p'_x)^2 + (\Del p'_y)^2 =
\gamma^2\,\frac{(h\,\nu)^2}{c_0^2}\,\left(\cos^2\vartheta -
2\,\beta\,\cos \vartheta + \beta^2 + \frac{\sin^2\vartheta}{\gamma^2}\right) =
\nonumber \\
\gamma^2\,\frac{(h\,\nu)^2}{c_0^2}\,[\cos^2\vartheta -
2\,\beta\,\cos \vartheta + \beta^2 + (1 - \beta^2)\sin^2\vartheta] =\nonumber \\
\gamma^2\,\frac{(h\,\nu)^2}{c_0^2}\,[1 -
2\,\beta\,\cos \vartheta + \beta^2\,(1 - \sin^2\vartheta)] =
\gamma^2\,\frac{(h\,\nu)^2}{c_0^2}\,(1 - \beta\,\cos \vartheta)^2
\label{Equation_10}
\end{eqnarray}
with a momentum change of
%
\begin{equation}
\Del P' = \gamma\,\frac{h\,\nu}{c_0}\,(1 - \beta\,\cos \vartheta)
\label{Equation_11}
\end{equation}
and with $\Del P' = h\,\nu'/c_0$ the Doppler formula is
%
\begin{equation}
\nu' = \gamma\,\nu\,(1 - \beta\,\cos \vartheta).
\label{Equation_12}
\end{equation}
The aberration formula then follows from Eqs.~(\ref{Equation_9})
and (\ref{Equation_11}) as
%
\begin{equation}
\cos \vartheta' = \frac{\Del p'_x}{\Del P'} =
\frac{\cos \vartheta - \beta}{1 - \beta\,\cos \vartheta}.
\label{Equation_13}
\end{equation}
Some special cases are:

There is no aberration in Fig.~\ref{fig:photon_1} for $\vartheta_0 = 0\deg$
and $\vartheta_4 = 180\deg$,
but, according to Eq.~(\ref{Equation_12}),
a longitudinal Doppler effects of
%
\begin{equation}
\nu_{\mp}' = \nu\,\gamma\,(1 \mp \beta) =
\nu\,\sqrt{\frac{1 \mp \beta}{1 \pm \beta}}
\label{Equation_14}
\end{equation}
will be observed.
On the other hand,
there is no Doppler effect for an emission angle $\theta_0$ defined in
system~S:
%
\begin{equation}
\cos \theta_0 = (\gamma - 1) / (\gamma \, \beta) \quad.
\label{Equation_15}
\end{equation}
However, $\theta_0$ changes to $\theta_0' = 180\deg - \theta_0$
in the system~S$'$ (cf. long green lines). \\
\citet{Hov98} demonstrated these relations in a three-dimensional treatment.
For $\theta' = 90\deg$ the transverse Doppler effect
%
\begin{equation}
\nu' = \nu/\gamma \quad.
\label{Equation_16}
\end{equation}
follows from Eqs.~(\ref{Equation_13}) and (\ref{Equation_12}).

The name ``transverse Doppler effect'' is a little misleading, because the
relativistic effect does not depend on the angles $\vartheta$ and $\vartheta'$,
and was, in fact, first observed with canal rays moving in both directions
along $\vartheta_0' = 0\deg$ and
$\vartheta_6' = 180\deg$ \citep{IveSti38,IveSti41}, and later with two-photon
spectroscopy in a similar geometry
\citep{Kaietal}; see also \citet{Saaetal}.

\section{Gravitational redshift}
\label{sec:Gravitation}

The discussion in this section will closely follow
our articles on the gravitational redshift \citep{WilDwi14,WilDwi19,WilDwi20}.

A relative wavelength
increase of $\approx 2 \times 10^{-6}$ was predicted for solar radiation by
\citet{Ein08} in 1908.
Experiments on Earth~\citep{PouReb,Craetal,Hayetal,KraLue,PouSni},
in space~\citep{BauWey} and in the Sun-Earth
system~\citep{StJ28,BlaRod,Bra63,Sni72,LoP91,Cacetal,TakUen} have
confirmed a relative frequency shift of
%
\begin{equation}
\frac{\nu' - \nu_0}{\nu_0} = \frac{\Del \nu}{\nu_0}
\approx \frac{U(r) - U_0}{c^2_0}~,
\label{Equation_17}
\end{equation}
where $\nu_0$ is the frequency of a certain transition at
the gravitational potential~$U_0$ and $\nu'$ is
the observed frequency there, if the emission caused by the same transition
had occurred at a potential~$U(r)$.
The potential at a distance~$r$ from a gravitational centre with
mass~$M$ is
%
\begin{equation}
U(r) = - \frac{G_{\rm N}\,M}{r}
\label{Equation_18}
\end{equation}
with $G_{\rm N} = 6.674 30 \times 10^{-11} \m^3\,\kg^{-1}\,\s^{-2}$
Newton's constant of gravity. The potential
is constraint in the weak-field approximation for non-relativistic
cases by \\$0 \le |U| \ll c^2_0$, cf., e.g., \citep{LanLif}.
To simplify the equations, we always put in this section
$U_0 = 0$ at $r = \infty$ and  $U(r) = U$.
In \citet{WilDwi19,WilDwi20}, we derived a good approximation of the
vacuum index of refraction as a function
of the distance~$r$ from mass~$M$:
%
\begin{equation}
\frac{1}{[n_{\rm G}(r)]} = \frac{c(r)}{c_0} \approx
1 - \frac{2\,G_{\rm N}\,M}{c_0^2\,r} = 1 + \frac{2\,U}{c_0^2}.
\label{Equation_19}
\end{equation}
The same index has been obtained, albeit with different
arguments, e.g., by \citet{Booetal,YeLin,Gupetal}.
The resulting approximation of the speed of light\footnote{An exact
expression is given in Eq.~(\ref{Equation_30}).}
\begin{equation}
c(r) \approx c_0\,\bigg(1 + \frac{2\,U}{c_0^2}\bigg)
\label{Equation_20}
\end{equation}
is in agreement with evaluations by
\citet{Sch60}, for a radial propagation\footnote{\citet{Ein12} states
explicitly that the speed at a certain location is not dependent on the
direction of the propagation.} in a central gravitational field, and
\citet{Oku00}\,--\,calculated on the basis of the standard Schwarzschild
metric. A decrease of the speed of light near the Sun, consistent with
Eq.~(\ref{Equation_20}), is not only supported by the predicted and
subsequently observed Shapiro delay
\citep{Sha64,Reaetal,Sha71,Kraetal,Baletal,KutZaj},
but also indirectly by the deflection of light
\citep{Ein16,Dys20}.

The question whether the redshift
occurs during the emission process or is a result of a
propagation effect is left open by \citet{Dic60} and \citet{Pou00}.
\citet{Okuetal} concluded that the energy of a propagating photon does not
change in a static gravitational field, however, momentum, velocity and
wavelength can change.
This conclusion is supported by \citet{Pet01} and \citet{Qua14}.
It is also consistent with the time
dilation of atomic clocks derived from the general theory of relativity
(GTR)~\citep{Ein16}.
\citet{Oha76} could not find a
loss of oscillations under steady-state conditions
in the Pound\,--\,Repka experiment.

Two statements by \citet{Ein08}, apparently in conflict,
are of interest in this context:

\begin{enumerate} \label{Spektrallinie0}
\item[ ]1. ``Since the oscillation process corresponding to a spectral line
can probably be seen as an intra-atomic process, the frequency of which
is determined by the ion alone, we can consider such an ion as a clock
with a certain frequency~~$\nu_0$.''
\end{enumerate}


and
\begin{enumerate}
\item[ ]2. ``The clock, therefore, runs more slowly, if it is positioned near
heavy masses. Consequently it follows that spectral lines of light
reaching us from the surface of large stars are displaced towards the
red end of the spectrum.''
\end{enumerate}


%
The first statement is probably correct,
if ``corresponding to a spectral line''
is neglected, since the electromagnetic forces acting in an atom
are many orders of magnitude larger than the gravitational forces.
The second statement is supported by many observations
in the Sun-Earth system, if the first sentence is not
included~\citep{StJ28,BlaRod,Bra63,Sni72,LoP91,Cacetal,TakUen}.

An easy solution to avoid this conflict is to postulate that the oscillating
atom, i.e. the `clock', does not necessarily have the same frequency
as the emitted spectral line \citep{WilDwi14}. In agreement with
\citet{Mueetal}, we will, therefore, consider the hypothesis
that no experiment is directly sensitive to the absolute potential~$U$.

Let us assume an atom~A with mass~$m$ and an energy
of~$E_0 = m\,c^2_0$ in the ground state located
at the gravitational potential~$U_0 = 0$.
With an energy difference~$\Del E_0$ from the ground
state to the excited atom~A$^*$, the mass in this state
is:
%
\begin{equation}
m + \Del m = \frac{1}{c^2_0}\,(E_0 + \Del E_0) = \frac{E^*_0}{c^2_0}~~,
\label{Equation_21}
\end{equation}
\citep{Ein05a,Lau20,Lau11}.
The rest energy of an excited atom at a distance~$r$ from
the centre of mass~$M$ is, cf. \citep{Oku00}:
%
\begin{equation}
E^* = E_0 + \Del E_0 + U\,(m + \Del m) ~,
\label{Equation_22}
\end{equation}
because lifting A* to $U_0 = 0$ would require the energy~$-U\,(m + \Del m)$.

The definition of the rest energy calls for some explanations:
If an excited atom with mass~$m + \Del m$ is lowered from
$U_0 = 0$ to~$U$,
its potential energy  will be converted, e.g., into kinetic energy
of the particle, $E_{\rm kin} = - U\,(m + \Del m)$. This energy
is provided by the gravitational field of the mass~$M$
interacting with the mass~$m + \Del m$ of the excited atom.
Since the atom is assumed to
be at rest at $U$, the kinetic energy is returned to the rest system of
mass~$M$ and the rest energy is $E^* = E_0 + \Del E_0 + U\,(m + \Del m)$.
If a photon had been emitted at $U$ with an
energy $h\,\nu = \Del E_0 = \Del m\,c_0^2$,
the energy of the atom would be $E = E_0 + U\,(m + \Del m)$.
To lift the mass $m$ to $U_0$ requires an energy of $- U\,m$ ,
leading to an energy of $E'_0 = E_0 - U\,\Del m$.
The photon, which does not change its energy during the
transit from $U$ to $U_0$,
could be converted to mass and re-excite the atom:
%
\begin{equation}
(E_0')^* = E_0 - U\,\Del m + \Del E_0.
\label{Equation_23}
\end{equation}
The different energies $(E_0')^* \ne E_0^*$ are in conflict
with energy conservation, cf. Eqs.~(\ref{Equation_21}) and (\ref{Equation_23}).
This can be avoided by either one of two assumptions:

1. The photon~$h\,\nu$ behaves in a gravitational field, contrarily
to the conclusions of, e.g., \citep{Okuetal,Pet01,Qua14}, the same way as a
massive body and would lose an energy of
$h\,\nu\,U/c_0^2 = (\Del m\,c_0^2)\,U/c_0^2 = \Del m\,U$.
The term $- U\,\Del m$ in Eq.~(\ref{Equation_23}) will then
not appear.

2. A photon would be emitted at $U$ with the rest energy of the transition
%
\begin{equation}
h\,\nu' = \Del m\,c_0^2 + U\,\Del m
\label{Equation_24}
\end{equation}
and travels with constant energy to $U_0 = 0$. We support this assumption,
but the question remains, how the gravitational potential~$U$ can be
determined.

Consider an excited atom\footnote{The atom should be part of a larger
mass in order to avoid significant kinetic energy effects,
cf. \citet{Moe58} .}~A$^*$
at~$U$: Assuming that it is neither able to sense the gravitational potential~$U$
nor the speed of light~$c(r)$ there, cf. Eqs.(\ref{Equation_20}) and
(\ref{Equation_30}).
The atomic `clock' is then driven by
%
\begin{equation}
\Del m\,c_0^2 = h\,\nu = |\vec{p_0}|\,c_0~,
\label{Equation_25}
\end{equation}
where the momentum vector~$\vec{p_0}$ of a photon emitted with $c_0$
has been included.
The emission of the photon by the atom assuming
energy and momentum conservation laws requires, however,
adjustments of the momentum and energy,
because $p_0\,c_0 \ne p_0\,c(r)$
would violate energy conservation. A momentum adjustment of $x$
according to equation
%
\begin{equation}
|p_0 - x|\,c_0 = |p_0 + x|\,c(r)
\label{Equation_26}
\end{equation}
is required. The emitted photon has a larger momentum, but a smaller energy
$-x\,c_0$ in line with the rest energy. The atom absorbes the
recoil momentum $-p_0 - x$, cf. Table~1 of Ref.~\citep{WilDwi14},
where the term `Interaction region' had been introduced.

The momentum adjustment~$x$ can be determined by solving the equation
%
\begin{equation}
\frac{p_0 - x}{p_0 + x} = \frac{c(r)}{c_0} \approx 1 + \frac{2\,U}{c_0^2}~,
\label{Equation_27}
\end{equation}
where the speed~$c(r)$ of Eq.~(\ref{Equation_20}) is interacting with the atom.
The adjustment is:
%
\begin{equation}
x \approx - p_0\,\frac{U}{c_0^2}~.
\label{Equation_28}
\end{equation}
The exact momentum~$x$ can also
be found by equating the emitted photon with the available energy
%
\begin{equation}
(p_0 - x)\,c_0  = \Del m\,c_0^2 + \Del m\,U~.
\label{Equation_29}
\end{equation}
Using Eq.~(\ref{Equation_26}), the speed~$c(r)$ can be obtained
after a lengthy calculation as:
%
\begin{equation}
c(r) = c_0\,\bigg(1 + \frac{2\,U}{c_0^2 - U}\bigg)~.
\label{Equation_30}
\end{equation}
For weak gravitational fields with $|U| \ll c_0^2$, this speed
agrees with the approximation in Eq.~(\ref{Equation_20}). The speed of
photons at $U$ thus plays an important r\^ole in regulating the
gravitational redshift.
The atom~A$^*$ at $U$ emits photons with an \\
energy~$h\,\nu' = \Del E_0 + U\,\Del m \approx p_0\,c_0\,(1 + U/c_0^2)$
in line with assumption~2 above and Eq.~(\ref{Equation_17})
as well as many observations.

Momentum considerations thus lead to
the result that only the rest energy of Eq.~(\ref{Equation_22})
can be emitted as photon and, therefore, assumption~2 is correct,
but it remains unclear, how the atom can determine the potential~$U$,
unless the speed of light at $U$ is involved.

We had derived the gravitational redshift
for electron-positron annihilations using the same method.
At $U_0 = 0$ we have electron and positron energies
of $E_{\rm e,p} = m_{\rm e,p}~c_0^2$ available for two
511~keV gamma photons.
If the particles fall towards the mass $M$, the potential energies are
converted to kinetic energie at the gravitatioal potential $U$:
%
\begin{equation}
E_{\rm e,p} = m_{\rm e,p}~c_0^2 + m_{\rm e,p}~U + E_{\rm kin}.
\label{Equation_31}
\end{equation}
Assuming the kinetic energies are absorbed there, the rest energies are
%
\begin{equation}
E^*_{\rm e,p} = m_{\rm e,p}~c_0^2 + m_{\rm e,p}~U.
\label{Equation_32}
\end{equation}
It would again be unclear how the emission process could sense the
potential~$U$ without energy and momentum conservation and a variable
speeds of light:
%
\begin{equation}
(p_0 - X)\,c_0 = (p_0 + X)\,c(r) = h\,\nu'.
\label{Equation_33}
\end{equation}
The momentum conservation is established by two gamma photons in
opposite directions with a momentum of $\pm (p_0 + X)$ and the expected
redshift of $h\,\nu' =  m_{\rm e,p}~c_0^2 + m_{\rm e,p}~U$.

\section{Cosmological redshift}
\label{sec:Cosmos}

In preparation for this section, it will be helpful to consider
the situation with \\ $|U(r)| \leq c_0^2$, e.g., at or near a black hole with
mass~$M$.
The Schwarzschild radius is:
%
\begin{equation}
r_{\rm S} = \frac{G_{\rm N}\,M}{c_0^2}~.
\label{Equation_34}
\end{equation}
The gravitational potential at $r_{\rm S}$ is:
%
\begin{equation}
U_{\rm S} = -\frac{G_{\rm N}\,M}{r_{\rm S}} = - c_0^2 ~.
\label{Equation_35}
\end{equation}
With Eq.~(\ref{Equation_30}) the speed of light
at the "event horizon" is $c_{\rm S} = 0$. Outside $r_{\rm S}$
the speed~$c$ increases with $r$.

One could ask the hypothetical question, what will happen inside the
black hole with decreasing $r$. As a black hole must be of spherical
shape, Newton's shell theorem \citep{New87} then tells us that only the mass
inside $r$ is relevant and the potential would
be:\footnote{Assuming a constant density, the mass of a sphere with
radius~$r < r_{\rm S}$ is proportional to $r^3$. The distance to the centre
is $r$ and thus the acceleration proportional to $r^{-2} \times r^3 = r$.
Since the acceleration is the gradient of the gravitational potential, $U(r)$
would be proportional to $r^2$.}
%
\begin{equation}
U(r) = U_{\rm S}\,\bigg(\frac{r}{r_{\rm S}}\bigg)^2 =
- c_0^2\,\bigg(\frac{r}{r_{\rm S}}\bigg)^2
\label{Equation_36}
\end{equation}
with $U_0 = 0$ at the centre.

We will not discuss the redshift of
close galaxies. If the small acceleration observed by \citet{PerSch} and
\citet{Rieetal} is
neglected, their redshift can reasonably well be explained by the
Doppler shift of moving sources, cf. \citep{Lem27,Hub29}, and conserves
energy and momentum.

On the other hand, distant galaxies, cf., e.g. \citep{KamRie},
and the Cosmic Microwave Background (CMB),
predicted by \citet{AlpHer} and discovered by
\citet{PenWil}, exhibits other redshift effects that have been observed
and extensively studied. As reference we want to mention the Nobel Lecture
by \citet{Smo07}, where further 84 references are listed, and
\citet{LamDol} with 67 more references.

The physical processes that led to the CMB radiation (CMBR) observations near
the Earth
at the present stage of the evolution of the Universe are differently
discussed in many
publications. In particular, the physics of the cosmic inflation is difficult
to understand,
cf., e.g. \citep{EarMos,VagLoe,Ijjetal}.
We will not consider it here.
There is, however, agreement on the observed temperature of the
radiation $T_0 = 2.726$~K \citep{Fix09}.

The black-body radiation\footnote{Cf., \citep{Ste79,Bol84,Wie93,Pla01}.}
is emitted after the deionization of the Big Bang (BB) plasma. This was
expected by \citet{Sil68} at a temperature of 4000~K, but present estimates are
near 3000~K. It is supposed to have happened 379\,000 years after the start of
the BB. The age of the Universe according to the \citet{Pla20} is
$13.787 \times 10^9$~years\footnote{\citet{Gup23} discussed 26.7 Gyr.} or
$4.3508 \times 10^{17}$\,s,
and it is difficult to understand how the CMBR
can now be seen at the Earth from all directions.

It is also very unclear how the
radiation after the emission as a 3000~K black-body spectrum is shifted
to one of 2.726~K. As outlined in Section~\ref{sec:Gravitation},
we agree with
\citet{Okuetal} and others that a photon does not change its
energy and frequency in a gravitational field, although the wavelength
will change with the gravitational potential.
We are, therefore, inclined to propose a completely different scenario
as follows:

The CMB dipole observations indicate a Solar System motion relative to
the cosmic rest frame of $369.82~\km~\s^{-1}$ \citep{Pla20}.
This can be explained
by a Doppler shift, see Section~\ref{sec:Doppler}: Eq.~(\ref{Equation_14})
and Fig.~\ref{fig:photon_1}.
The small value could mean that the
position of our {\em future} Galaxy
would be relatively close to the centre of the BB, whereas
the outer parts with most of the matter, energy and probably antimatter
would be propagating outwards with a peak velocity of $v_{\rm exp} \approx c_0$.

At the beginning of this section, the gravitational potential inside a black
hole was considered, cf. Eq.~(\ref{Equation_36}).
This might be helpful for the problem at hand: Are we near the centre
of a black hole (BH)? (cf., e.g., \citep{Pat72,SmoTem,Pop16,Par23}).

Shortly before the deionization, the photons interact with the plasma
with a 3000~K black-body spectrum. At deionization, the photons
emitted in the direction to the centre of the BH, i.e. perpendicular to
the plasma sphere, will experience a Doppler shift. Since this effect
depends only on the relative velocities of source and receiver, we
can find with Eq.~(\ref{Equation_14}) and $\theta_0 = 0\deg$
the required $\beta = v_{\rm exp}/c_0$ for a
redshift of $z = 1100$: $\beta = 0.999\,998\,35$.
This process, as outlined in Section~2, conserves energy and momentum.
\begin{figure}[H]
\centering
\includegraphics[width=\textwidth]{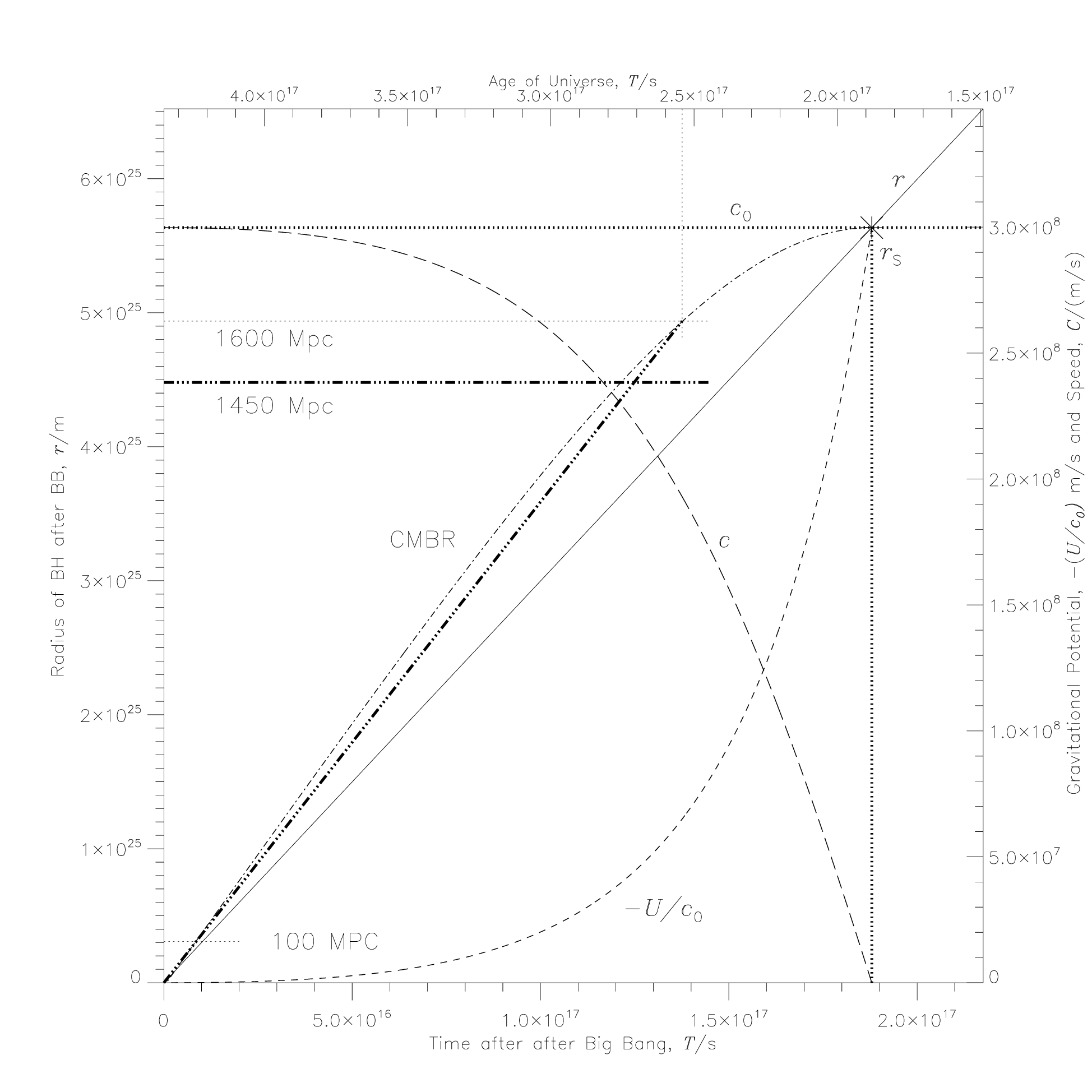}
\caption[]{\small
In this diagram we first want to show the parameters of the Black Hole (BH)
after the Big Bang (BB)
in our scenario. $r$ (solid line) is the radius with respect to the time scale on the lower
axis and the scale on the left assuming a peak expansion rate of $c_0$.
The Schwarzschild radius $r_{\rm S} = 5.6348 \times 10^{25}$\,m is reached
after $1.8796 \times 10^{17}$\,s, i.e. 0.432 of the age of the Universe
(see calculation in the text). Secondly, the diagram displays the
gravitational potential ($-U/c_0$ shown as short dashed line with a scale on the
right and a reverse time scale in the
upper axis), which leads to a variation of the speed
of light $c$ (long dashed line) according to Eq.~(\ref{Equation_30}).
This speed determines the arrival
of the CMBR (dash-dot line) at the origin at the present age of the Universe,
i.e. now. To demonstrate a hypothetical explanation of the Hubble tension,
the radius at 1600~Mpc is selected as an example.
The dash-dot-dot-dot line from 1600~Mpc to the origin between
$T = 2.54 \times 10^{17}$\,s and $4.3508 \times 10^{17}$\,s
of the upper axis gives a
mean speed of light of $2.72 \times 10^8\,\m~\s^{-1}$.
Observations made under the wrong assumption of a constant speed of light~$c_0$
refer to 1450~Mpc and not to 1600~Mpc.
Some additional lines
are added for orientation. More details are given in the text.}
\label{fig:photon_2}
\end{figure}
Since $\beta$ is so large,
$|U|$ will be close to $c_0^2$ at $r_{\rm S}$ in Eq.~(\ref{Equation_36})
and decrease with $r^2$ as the photons move towards the centre.
As the Universe further expands with $c_0$, we also assume
an increase of $r_{\rm S}(T)$ with time~$T$. This has no direct effect
on the gravitational
potential, because of the shell theorem, but will lead to a decrease of
the mass density inside the BH. A good approximation of the additional decrease of $U$
might therefore be $[r_{\rm S}/r_{\rm S}(T)]^3$.

We now have to understand, how the CMBR reaches the
Earth from the starting point at $r_{\rm S}$ taking into account
the dependence of the photon speed on the gravitational potential, cf.
Eq.~(\ref{Equation_30}).
We will assume for $r_{\rm S}$
a variable fraction~$X$ of the present age of the Universe and integrate
the path $C\,{\rm d}t$ in time steps of $1/10^4$ of the age of the
Universe,
${\rm d}t = 4.3508 \times 10^{13}$\,s,
until the photons arrive at the centre of the BH at the present time.
We find a fraction of $X = 0.432$, i.e. a travel time on the upper axis of
($4.3508 - 1.8796) \times 10^{17}$\,s. The CMBR reaches us from a distance of
$r_{\rm S} = 5.6348\times 10^{25}$\,m. This is shown by the dashed-dotted
line in Fig.~\ref{fig:photon_2} together with $-U/c_0$
as short-dashed line.

The variable speed of light suggested by this scenario
might also give a hint to explain the Hubble tension, cf., e.g.
\citep{KamRie,Pouetal}, with observed values between about
$67.4~\km~\s^{-1}\,\Mpc^{-1}$ for remote and
$73~\km~\s^{-1}\,\Mpc^{-1}$ for closer astronomical objects.
Considering two examples at 1600~Mpc and 100~Mpc, we find in the first
case a mean speed of $2.72 \times 10^8\,\m~\s^{-1}$ and in the second case near the
origin $2.997 \times 10^8\,\m~\s^{-1}$, i.e. very close to $c_0$.

The observations and evaluations leading to the Hubble tension
are of many different types,
e.g., \citep{Phietal,Hut23,Cametal,Veretal,Sco25},
but it appears as if the speed of light was assumed to be constant at $c_0$
in all cases.

The travel time to 1600~Mpc with $c_0$ is $t_0 = 1600~\Mpc/c_0$.
With the reduced speed a distance of
$t_0 \times 2.72 \times 10^8\,\m~\s^{-1} = 1450~\Mpc$ would be reached.
Observers assuming a constant speed of light~$c_0$ and measuring
$H_0 = 67.4~\km~\s^{-1}\,\Mpc^{-1}$ there, think that they have
obtained this value for 1600~Mpc and divide the measured
velocities of the astronomical objects by 1600 and not by 1450~Mpc.
A correction of this error gives:
\begin{equation}
67.4~\km~\s^{-1}\,\Mpc^{-1} \times \frac{1600}{1450} =
74.2~\km~\s^{-1}\,\Mpc^{-1} \quad .
\label{Equation_37}
\end{equation}
No such correction is required for nearer objects, because the speed of light
is close to $c_0$.

A more detailed study could extend the range to larger distances and would also
have to consider the effect of gravitational redshifts,
cf. Section~\ref{sec:Gravitation}. The proposed solution of the Hubble
tension would lead to a constant $H_0$.
The expansion of the universe at its front is assumed to proceed with the
speed of light~$c_0$. The age of the universe is expected to be
$T_0 = 4.3508 \times 10^{17}$\,s, leading to a radius of
$r_0 = 1.3044 \times 10^{26}$\,m or 4227~Mpc and a Hubble constant of
$H_0~=~70.92~\km~\s^{-1}\,\Mpc^{-1}$.

\section{Conclusions}
\label{sec:Cons}
The fact that \citet{OrtIba} grounded the redshift effects only on the
principle of energy conservation induced us to reiterate that both
Einstein and Fermi, among others, involve not only energy, but
also momentum conservation. The discussion of the cosmological redshift
led to the hypothetical scenario that the surrounding Universe is
situated inside an expanding black hole. The variable speed of light
might provide an explanation of the Hubble tension.

We contend that an energy-dependent unified redshift relation should also
incorporate momentum in order to apply
it for a wide variety of systems, whether quantum, classical, or relativistic.


\end{document}